\documentstyle[12pt,epsfig,rotating]{article}
\newcommand{\AmS}{{\protect\the\textfont2
  A\kern-.1667em\lower.5ex\hbox{M}\kern-.125emS}}
    \def\r#1{$^{[#1]}$}

\def\ifmath#1{\relax\ifmmode #1\else $#1$\fi}%

\def\s2{\hskip2pt}  \def\d{{\rm d}}    \def\e{\rm e}
\def\f{\left}   \def\g{\right}
\newcommand{\beqa}{\begin{eqnarray}} \newcommand{\eeqa}{\end{eqnarray}  }
\newcommand{\beqan}{\begin{eqnarray*}} \newcommand{\eeqan}{\end{eqnarray*}}
\newcommand{\beq}{\begin{equation}} \newcommand{\eeq}{\end{equation}  }
\def\cl{\centerline} \def\bcc{\begin {center}} \def\ecc{\end {center}}
\def\btbl{\begin{tabular}} \def\etbl{\end{tabular}}

\begin{document}
\null{}\vskip -1.2cm
\hskip12cm{\bf HZPP-0002}

\hskip12cm Feb. 20, 2000

\vskip1cm

\centerline{\Large  Thermal Equilibration and Ellipsoidal Expansion}
\vskip0.3cm
\centerline{\Large of Rotationally-Symmetrical Longitudinal Flow }
\vskip0.3cm
\centerline{\Large in Relativistc Heavy Ion Collisions\footnote{Work
supported in part by the NSFC under project 19775018.}}
\vskip1cm
\centerline{\large Feng Shengqin, Liu Feng and Liu Lianshou}
\vskip0.4cm
\centerline{\small Institute of Particle Physics, Huazhong Normal University, 
Wuhan, 430079, China}

\vskip0.9cm

\vskip2cm
\begin{center}{\large ABSTRACT}\end{center}
\vskip0.5cm
\begin{center}\begin{minipage}{124mm}
{\small \hskip0.8cm
A rotationally-symmetrical ellipsoidal flow model is proposed for the 
relativistic heavy-ion collisions and compared with the 14.6 A GeV/$c$
Si-Al and 10.8 A GeV/$c$ Au-Au collision data. The large stopping in the
heavier collision system and heavier produced particles is accounted 
for by using the  ellipsoidal flow picture.  The central dip in the proton 
and deuteron rapidity distributions for Si-Al collision are reproduced.}
\end{minipage}\end{center}

\newpage

\noindent
{\bf I. Introduction}\\
\noindent

The experimental finding that colliding nuclei are not transparent but undergo 
a violent reaction in central collisions represents one of the major 
motivations for the study of ultra-relativistic heavy ion collisions at the 
CERN/SPS, BNL/AGS and also at the future BNL/RICH and CERN/LHC. Of central 
importance is the ability of understanding to what extent the nuclear
matter has been compressed and heated.

The study of collective flow in high energy nuclear collisions has attracted 
increasing attention from both experimental\r{1} and theoretical\r{2}
point of view. 
The rich physics of longitudinal and transverse flow is due to their
sensitivity to the system evolution at early time. The expansion and cooling 
of the heated and highly compressed matter could lead to considerable 
collectivity in the final state. Due to the high pressure, particles might
be boosted in the transverse and longitudinal directions. The collective 
expansion of the system created during a heavy-ion collision implies 
space-momentum correlation in particle distributions at freeze-out. 

The experimental data of the rapidity distributions of produced particles
in 14.6 A GeV/$c$ Si-Al collisions have been ultilized to study the 
collective expansion using a cylindrically-symmetrical flow model\r{3,4}.
The model results fit well with the experimental distribution of pion, 
but is too narrow in the case of heavier particles proton and deuteron. 
In particular, the central dip, which can be clearly seen in the distribution 
of proton, is failed to be reproduced. 

More recently, E877 Collaboration\r{5} has published their data for 
10.8 A GeV/$c$ Au-Au collisions, which provide a good chance to compare 
the stopping power in the collision systems of different sizes. 
The possible central peak of the rapidity distribution of proton at
around mid-rapidity, which was obtained through extrapolating the 
experimental data to mid-rapidity using RQMD model\r{6}, has been taken as an 
evidence for the increasing of stopping power, but the reliability of this
extrapolation is model-dependent.

It has been shown earlier\r{7} that the ellipsoidal expansion is a simple 
way to take the nuclear stopping into account.  In the present paper we 
propose a  rotationally-symmetrical ellipsoidal flow model
to describe the space-time evolusion in relativistic heavy ion collisions.
The large stopping in the heavier collision system and heavier produced 
particles is described by using this picture. The central dips in the 
proton and deuteron rapidity distributions for Si-Al collisions are 
reproduced.

In section II the rotationally-symmetrical ellipsoidal flow model is 
formulated. The results of the model are given and compared with
the experimental data in section III.  A short summary and conclusions are 
given in section IV. In order to avoid the complexity in the production
of strange particles and concentrate on the expansion of the system, we will 
discuss in this paper only normal non-strange particles ------ pions, protons 
and deuterons.

\newpage
\vskip0.2cm
\noindent
{\bf II. Rotationally-symmetrical ellipsoidal flow  } \\
\vskip0.2cm

Firstly, let us briefly recall the fireball scenario of relativistic heavy ion 
collisions. 

Since the temperature at freeze-out exceeds 100 MeV, the 
Boltzmann approximation is used. Transformed into rapidity $y$ and
transverse momentum $p_{t}$ this implies\r{4}:
\begin{equation}  
  E\frac {\d^{3}n}{\d^{3}p} \propto E\e^{(-E/T)}=    
  m_{t}\cosh(y) \e^{(-m_{t}\cosh(y)/T)} 
\end{equation}
Here $m_{t}=\sqrt{m^{2}+p_{t}^{2}}$ is the transverse mass, $m$ is the mass 
of the produced particles at freeze-out.

The rapidity is defined as $y=\tanh^{-1}(p_{l}/E)$, where $p_{l}$ is the 
longitudinal momentum of the produced particle. Substituting into Eq.(1) 
and integrating over $m_{t}$, we get the rapidity distribution of the 
isotropic thermal source,
\begin{equation}  
\frac {\d n_{\rm iso}}{\d y} \propto 
\frac {m^{2}T}{(2\pi)^{2}}(1+2\xi_0+2\xi_0^{2})\e^{(-1/\xi_0)}.  
\end{equation}
Here $\xi_0={T}/{m\cosh(y)}$. 

However, the momentum distribution of the measured particles is certainly
not isotropic. It is privileged in the direction of the incident nuclei. 
This is because the produced hadrons still carry
their parent's kinematic information, making the longitudinal 
direction more populated than the transverse ones.
The simplest way\r{3,4} to account for this anisotropy is to add the 
contribution from a set of fire-balls, sketched schematically in Fig.1 as 
dashed circles, with centers located uniformly in the rapidity region 
[$-y_0, y_0$]. The corresponding rapidity distribution is obtained through 
changing the $\xi_0$ in Eq,(2) into $\xi={T}/{m\cosh(y-y')}$ and integrating 
over $y'$ from $-y_0$ to $y_0$:
\begin{equation}  
\frac {\d n_{\rm cyl}}{\d y} \propto 
\int_{-y_0}^{y_0} \d y' 
\frac {m^{2}T}{(2\pi)^{2}}(1+2\xi+2\xi^{2})\e^{(-1/\xi)},   
\end{equation}
$\xi={T}/{m\cosh(y-y')}$. Equivalently, we can also use the angular variable 
$\Theta$ defined by $\Theta = 2 \tan^{-1} \exp (-y')$,    
and change the integration 
variable in Eq.(3) to $\Theta$, cf. the solid circle and lines in Fig.1.

This simple approach fits the rapidity distribution of pions well but
failed to reproduce the central dip in heavier produced particles, which
is clearly seen in the experimental distribution of protons and has some
evidence in the distribution of deuterons.

Note that in this model the longitudinal and transverse expansions of the 
system are totally independent. This is a crude approximation. A more 
reasonable picture is an ellipsoidal expansion.  For simplicity the rotational 
symmetry arround the longitudinal direction is still assumed, but the 
emission angle is now 
\beq   
\theta = \tan^{-1} (e \tan\Theta),
\eeq
where $e$ ($0\leq e \leq 1$) is the ellipticity, cf. Fig.2.

In this model the ellipticity parameter $e$
represents the degree of anisotropy of flow in the transverse and longitudinal 
direction. The smaller is $e$, the more anisotropic is the flow.
The nuclear stopping can be taken into account in this way.

Subsituting Eq.(4) together with $y'_{\rm e}=-\ln\s2\tan(\theta/2)$ into 
Eq.(3), the rapidity distribution is obtained:
\beqa   
\frac {\d n_{\rm ellip}}{\d y} &=& eKm^2T 
\int_{\theta_{\rm min}}^{\theta_{\rm max}} 
\f(1+\frac{2T}{m\cosh(y-y'_{\rm e})}+ \frac{2T^2}{m^2\cosh^2(y-y'_{\rm e})} \g) 
\nonumber \\
&\null{}& \hskip6cm \times
\exp(-m\cosh(y-y'_{\rm e}) / T)    
Q(\theta) \d \theta,
\eeqa
\beq 
y'_{\rm e}=-\ln\s2\tan(\theta/2)\ , \qquad 
Q(\theta) = \frac{1}{\sqrt{e^2+\tan^2\theta} |\cos\theta| \sin\theta}.
\eeq
\noindent
Here $\theta_{\rm min}=2\tan^{-1}(e^{-y'_{\rm e0}})$, 
$\theta_{\rm max}=2\tan^{-1}(e^{y'_{\rm e0}})$.
$y'_{\rm e0}$ is the rapidity limit which confines the 
rapidity interval of ellipsoidal flow. We
treat it together with the ellipticity $e$ as two free papameters
of the model to fit the amount of flow and stopping required by 
the data.  

\vskip1.2cm
\noindent
{\bf III. Comparison with experiments  } \\
\vskip0.2cm
The rapidity distributions of pion, proton and deuteron 
for 14.6 A GeV/$c$ Si-Al collisions{\r{9,10}}, 
are given in Fig.3 ($a, b$ and $c)$. 
The dashed, dotted and solid lines correspond to the results from 
isotropical thermal model, cylindrically-symmetrical flow model and 
rotationally-symmetrical ellipsoidal flow (RSEF) model 
respectively. The rapidity limit $y'_{\rm e0}$ and the ellipticity $e$
used in the calculation are listed in Table I. The  rapidity limit 
$y'_{\rm 0}$ used in the cylindrically-symmetrical flow model of
Ref. [4] is also listed for comparison.
The parameter $T$ is chosen to be 0.12 GeV following 
Ref.[4].

\vskip0.5cm
\cl{Table I \ \ The value of model-parameters}
\vskip-0.5cm
\bcc\btbl{|c|c|c|c|c|c|}\hline
           & \multicolumn{3}{|c|}{Si-Al Collisions}
        &\multicolumn{2}{|c|}{Au-Au Collisions} \\ \cline{2-6}
 Parameter & $\pi$ & p & d  & $\pi$ & p  
\\ \hline
$e$ & 0.28 & 0.52 & 0.56 & 0.32 & 0.58 
\\ \hline
$y'_{\rm e0}$ & 1.35 & 1.35 & 1.35 & 1.05& 1.05  
\\ \hline
$y'_{\rm 0}$ & 1.15 & 1.15 & 1.15 & &   \\
\hline
\etbl\ecc

It can be seen from the figures  that
the RSEF model reproduces the central dip of rapidity distribution 
of heavier particles  (proton and deuteron) in coincidence with 
the experimental findings, while for light particles (pions) there is 
a plateau instead of dip at central rapidity. 
Note that the appearance or disappearance of central dip is insensitive to
the rapidity limit $y'_{\rm e0}$ but depends strongly on the magnitude 
of the ellipicity $e$ and the mass $m$ of the produced particles.  
For the heavier particles (proton and deuteron) a central dip appear for
$e < 0.8$, but for light particles  (pions) there is no dip even when $e$ 
is as small as 0.28, cf. Table I and Fig. 3.

It can also be seen from Table I that $e_{\rm d} > e_{\rm p} > e_\pi$.
It means that the system is less alongated for proton and deuteron than for 
pion.  This describes nuclear stopping. 

On the other hand, the width of the rapidity distributions are mianly
controlled by the parameter $y'_{\rm e0}$. The value $y'_{\rm e0}=1.35$,
a little bigger than 1.15 used in the cylindrically-symmetrical flow model of
Ref. [4] can account for the wide distriution of heavier particles (protons
and deuterons) and at the same time fits the pion-distribution well.

In Fig.4 are shown the rapidity distributions of pions and protons 
for Au+Au collisions at 10.8 A GeV/$c${\r 5}.
The solid and dashed lines correspond to
the results of RSEF model (with parameters listed in Table I)
and cylindrically-symmetrical flow model respectively. 
The latter are obtained also using RSEF with the same rapidity limit $y'_0$ 
as the $y'_{\rm e0}$ listed in Table I but with ellipicity $e=1$.
The histogram is the result from the RQMD model.

It can be seen from Fig.4 that in the RSEF model there is a shallow dip
(plateau) in the central rapidity of the distribution of proton, instead of 
a central peak as predicted by the cylindrically-symmetrical flow model. 
However, the presently available experimental data are restricted to the large
rapidity. The peak at central rapidity is the extrapolation of data using
RQMD and is model dependent. 
It is intersting to see whether the prediction of a central dip
(plateau) or a central peak will be observed in future experiments.

Comparing the parameter values for Si-Al (smaller colliding nuclei) and 
Au-Au (larger colliding nuclei) collisions listed in Table I, it can be seen 
that the rapidity limit $y'_{\rm e0}$ is smaller and the elliticity $e$ is 
bigger for the larger colliding nuclei than for the smaller ones. Both of
these two show that the hadronic system formed from the larger colliding 
nuclei is less alongated, i.e. there is stronger nuclear stopping in the
collision of larger nuclei.

\vskip1.2cm
\noindent
{\bf IV. Summary and Conclusions}
\vskip0.2cm

In high energy heavy-ion collisions, due to the transparency of the nucleus 
the participants
will not lose the historical vestiges and the produced hadrons will 
carry some of their 
parent's memory of motion, leading to the unequivalence in longitudinal
and transverse directions. So it is reasonable to assume that the flow of 
produced particle is privileged
in the longitudinal direction. This picture has been used by lots of 
models\r{3,8}. Here we should mention two 
thermal and hydrodynamic models, one is the the 
boost-invariant longitudinal expansion model postulated by Bjorken \r{8}
which can explain such an anisotropy already at the level of 
particle production.
This model has been
formulated for asymptotically high energies, where the rapidity 
distribution of
produced particles establishes a plateau at midrapidity. The second model 
is the 
cylindrical symmetry flow model postulated first by Schnedermann, 
Sollfrank and Heinz\r{3} which
account for the anisotropy of longitudinal and transverse direction 
by adding the contribution from
a set of fire-balls with centers located uniformly in the rapidity 
region [-$y'_{\rm 0}$,$y'_{\rm 0}$] in the longitudinal direction,
sketched schematically in Fig.1 as dasheded circles. 
In this model the centers of fire-balls distribute uniformly in rapidity, 
and so it gives the picture that longitudinal and transverse expansion are 
totally independent.  It can account for the wider rapidity distribution
when comparing to  the prediction of the pure thermal 
isotropically model but failed to reproduce
the central dip in the proton and deuteron rapidity distributions.  

In this paper, we propose an ellipsoidal expansion model with 
rotationally-symmetrical longitudinal flow (rotationally-symmetrical
ellipsoidal flow model, RSEF)
which realizes that the centers of fire-balls are distributed un-uniformly
in a rotationally-symmetrical ellipsoidal shape around the longitudinal 
direction. The ellipticity parameter $e$ can account for
the extent of anisotropy of phase space in transverse 
and longitudinal expansion. The central dip in the proton and deuteron 
rapidity distributions and the central peak in the pion distribution are 
well reproduced simultaneously from this model. 

It is found that the depth of the central dip of the heavier particle 
distributions depends strongly on the magnitude of the ellipicity $e$. 
In other words, the anisotropy in transverse and 
longitudinal directions of the ellipsiod of phase space, which is given by 
the ellipticity $e$, determines also the depth of the central dip for heavier 
particles.  

Through comparing the feature of collision systems of different size, 
we found that the maximum 
flow velocities are larger for the lighter collision systems than the 
heavier ones, which suggests, together with smaller $e$,
a larger stopping in the larger collision system. 


\vskip2cm
\noindent{\bf\large\bf Figure captions}

\vskip0.8cm
\noindent {\large Fig.1} \ Schematic sketch of the 
cylindrically-symmetrical flow model.

\vskip0.8cm
\noindent {\large Fig.2} \ Schematic sketch of the 
Rotationally-symmetrical ellipsoidal flow model

\vskip0.8cm
\noindent {\large Fig.3} \ Rapidity distributions for central 14.6 A GeV/$c$
Si+Al collisions. Open circles --- data from Si+Al collisions [8,9]; 
Dashed lines --- isotropical thermal model; Dotted lines ---
cylindrically-symmetrical flow model; solid  line --- rotationally-symmetrical
ellipsoidal flow (RSEF) model. Temperature  $T=0.12 GeV$. 
Fig.3 ($a$), ($b$) and ($c$) are for the pion, 
proton and deuteron distributions respectively. 
\vskip0.8cm
\noindent
{\large Fig.4} \ Rapidity distributions for pions and protons in 
central Au+Au collisions with 10.8  A GeV/$c$. Full circles 
represent measured data, open circles reflected data. 
The solid  line is our calculation using the RSEF model. 
The histogram shows the results from RQMD calculations and the dotted line
is the results from the prediction of cylinderically-symmetrical 
flow model. The temperature $T=0.14GeV$. Fig.4 ($a$)
and Fig.4 ($b$) are for the pion and proton distributions respectively.

\newpage
\begin{center}
\begin{picture}(250,450)
\put(-40,340)
{
{\epsfig{file=fig1.epsi,width=350pt,height=150pt}}
}
\end{picture}
\end{center}

\vskip-12cm
\cl{Fig. 1}

\begin{center}
\begin{picture}(250,450)
\put(30,200)
{
{\epsfig{file=fig2.epsi,width=220pt,height=170pt}}
}
\end{picture}
\end{center}
\vskip-6cm
\cl{Fig. 2}

\newpage
\begin{center}
\begin{picture}(250,450)
\put(-15,100)
{
{\epsfig{file=fig3.epsi,width=240pt,height=380pt}}
}
\end{picture}
\end{center}

\vskip-2cm
\cl{Fig. 3}

\newpage
\begin{center}
\begin{picture}(250,450)
\put(0,200)
{
{\epsfig{file=fig4.epsi,width=220pt,height=240pt}}
}
\end{picture}
\end{center}
\vskip-6cm
\cl{Fig. 4}

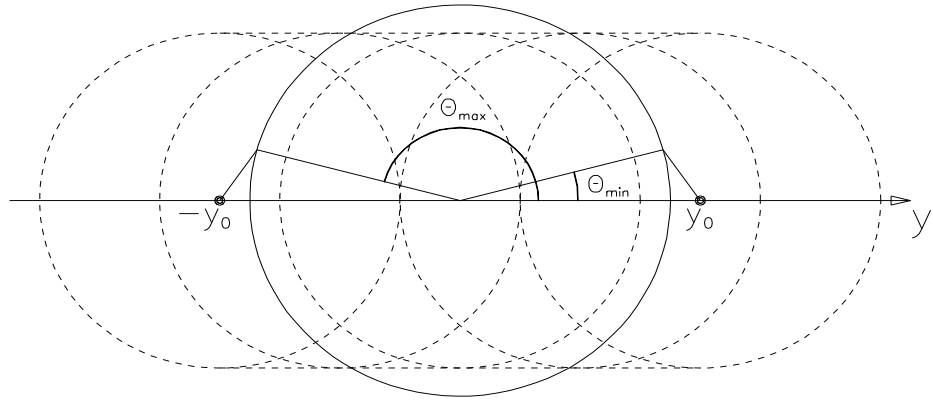
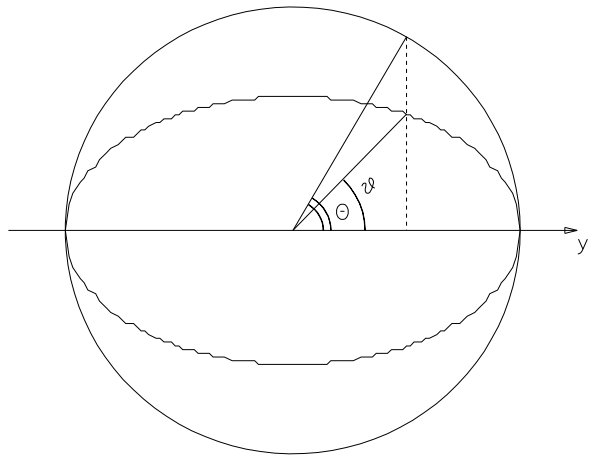
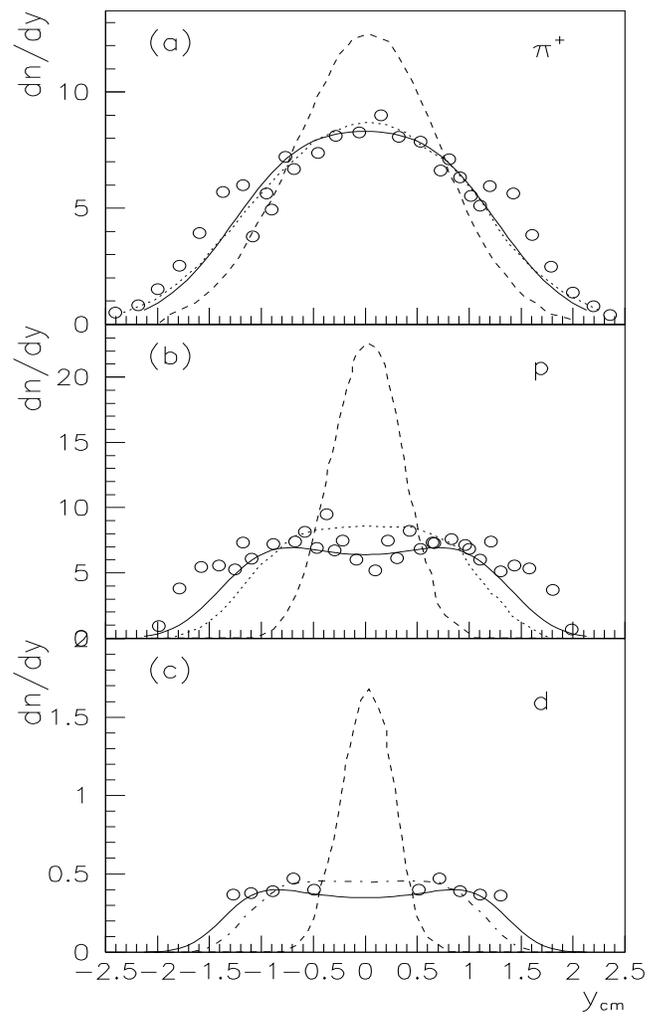
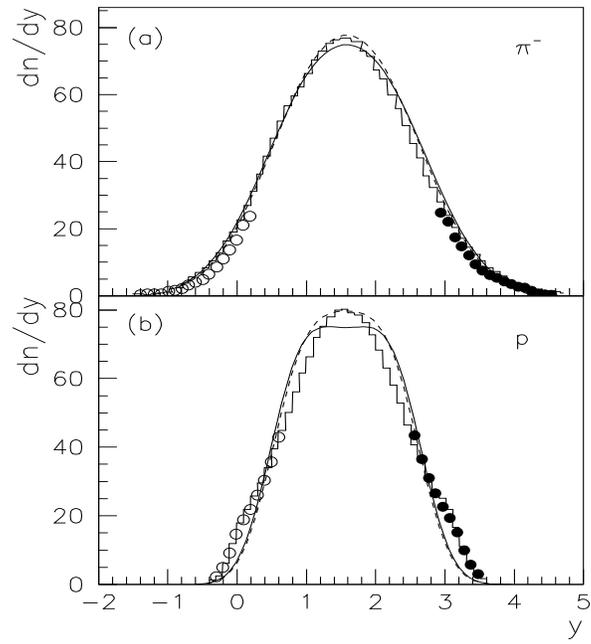
\end{document}